\documentclass{article}
\usepackage[english]{babel}
\usepackage{graphicx,graphics}

\newcommand{\ket}[1]{\ensuremath {|\: #1 \: \rangle}}
\newcommand{\bra}[1]{\ensuremath{\langle \: #1 \:|}}
\newcommand{\braket}[2]{\ensuremath{\langle \: #1 \: | \: #2 \: \rangle}}
\newcommand{\ketbra}[2]{\ensuremath{| \: #1 \:\rangle \langle \: #2 \:  |}}
\newcommand{\ves}[2]{\ensuremath{#1_1,#1_2, \ldots, #1_{#2}}}

\begin{document}
\title{Entropy generation in a model of reversible computation.}
%
%
\author{
\small{Diego de Falco}\\
\small{Dipartimento di Scienze dell'Informazione}\\
\small{Universit\`a degli Studi di Milano}\\
\small{via Comelico 39, 20135 Milano, Italy}\\
\small{\itshape{e-mail: defalco@dsi.unimi.it}}
\and
\small{Dario Tamascelli}\\
\small{Dipartimento di Matematica}\\
\small{Universit\`a degli Studi di Milano}\\
\small{via Saldini 50, 20133 Milano, Italy}\\
\small{\itshape{e-mail: tama@mat.unimi.it}}}
\date{}
\maketitle%

\begin{abstract}
We present a model in which, due to the quantum nature of the signals controlling the implementation time of successive unitary computational steps, \emph{physical} irreversibility appears in the execution of a \emph{logically} reversible computation.
\end{abstract}
\section*{Introduction}
Quantum computation is mostly seen as a static process: we are given an algorithm $A$ that must be decomposed in elementary computational steps which have to be unitary operators  acting on the state space of the input/output register; the computing process consists then in the sequential application of the computational primitives to the register. This approach is the same adopted in classic computer science and has its genuine motivation in the functioning of a Turing machine in its simplest version: at \emph{each time step} the tape is read and an operation is performed. In our desktops the time steps correspond to the ticks of the internal clock of the processor which ``beats the bongos'' at a given frequency.\\
As soon as we leave the computational paradigm to come down to the implementation of quantum algorithms, it turns out that the computational process \emph{is} the \emph{time-evolution} of a physical system prepared and acted upon by external agents in such a way that the initial state $\ket{R(1)}$ of the register evolves, at a certain time $\bar{t}$, into the final state $A\ket{R(1)}$ obtained as 
\begin{eqnarray}
	 A \ket{R(1)}& = & U_{s-1}U_{s-2}\ldots U_2 U_1 \ket{R(1)} =\nonumber \\
	& = & \exp(-i H_{s-1}\tau_{s-1}) \exp(-iH_{s-2}\tau_{s-2})\ldots \nonumber\\
	& & \ldots \exp(-iH_{2}\tau_2)\exp(-iH_{1}\tau_1)\ket{R(1)} \nonumber
\end{eqnarray}
$H_i,\ i=1,\ldots,s-1$, being suitable Hamiltonian operators. This kind of external control (switching on and off the Hamiltonians at given times) is a physically very hard task to perform. The alternative definition of a single time independent Hamiltonian $K$ under which the register can autonomously evolve from \ket{R(1)} to $A\ket{R(1)}=\exp(-iK\bar{t})\ket{R(1)}$ turns out to be an equivalently difficult problem (see, for example, \cite{peres85}).\\
Feynman's proposal for a quantum computer \cite{feyn86} represents a way out from this: by coupling the input/output register to additional degrees of freedom \emph{acting as a clock}, it becomes immediate to define a single time independent Hamiltonian determining the desired evolution. Furthermore, being able to implement the Toffoli gate, the computing model is able to compute all the function computable by a deterministic reversible Turing machine.\\
Some of the features of the model have already been investigated in previous work \cite{apolloni02}\cite{defa03}\cite{defa04}. In particular, in \cite{defa03} we have shown that the use of conditional jumps makes Feynman's computer an implementation of the continuous time quantum walk computational paradigm \cite{farhi03}.\\
In this work we analyze the coupling between the input/output register and the clocking mechanism and the appearance of physical irreversibility in the context of logically reversible computation. 
\section*{Feynman's cursor model}
Let $\rho(1)=\ketbra{R(1)}{R(1)}$, with $\ket{R(1)} \in \mathcal{H}_{register}$, be the initial state of the \emph{input/output register}. Let \ves{U}{s-1} be the unitary operators representing the successive ``primitive'' steps of the computation to be performed.\\
Set
\begin{equation}
	\ket{R(x)}= U_{x-1}\ \ket{R(x-1)},\ 1 < x \leq s.
\end{equation}
Call $\mathcal{K}_r$ the subspace of $\mathcal{H}_{register}$ spanned by the vectors $\ket{R(1)}, \ket{R(2)},\ldots,\ket{R(s)}$.\\
Set
\begin{equation}
	d=dim(\mathcal{K}_r).
\end{equation}
Following the approach of \cite{feyn86}, we model the clocking mechanism, which sequentially applies the transformations \ves{U}{s-1} to the register, with a quantum mechanical system, the \emph{cursor}.\\
We call $\mathcal{H}_{cursor}$ the $s-$dimensional state space of this system and refer it to a selected orthonormal basis $\ket{C(1)},\ket{C(2)},\ldots,\ket{C(s)}$.\\
We call \emph{position of the cursor} the observable $Q$ acting on the vectors of this basis as
\begin{equation}
	Q \ket{C(x)}= x\ \ket{C(x)},\ 1 \leq x \leq s.
\end{equation}
We suppose that the state of the overall system, the \emph{machine}, evolves in the Hilbert space $\mathcal{H}_{machine}= \mathcal{H}_{register} \otimes \mathcal{H}_{cursor}$ under the action of a Hamiltonian of the form
\begin{equation} \label{eq:hamiltonian}
	H= -\frac{\lambda}{2} \sum_{x=1}^{s-1} U_x \otimes \ketbra{C(x+1)}{C(x)}+ U^{-1}_x \otimes \ketbra{C(x)}{C(x+1)}.
\end{equation}
We suppose, furthermore, that the state of the machine is, at time $t=0$, represented by the vector
\begin{equation} \label{eq:condiniziali}
	\ket{M(0)}= \ket{R(1)} \otimes \ket{C(1)}.
\end{equation}
It is well known \cite{gramss95} that for every time $t$ it is then
\begin{equation}
	\ket{M(t)}=\sum_{x=1}^s c(t,x;s) \ \ket{R(x)} \otimes \ket{C(x)}
\end{equation}
where
\begin{equation}
	c(t,x;s)= \frac{2}{s+1} \sum_{k=1}^s \exp\left[ i \lambda t  \cos(\vartheta(k;s))\right] \sin(\vartheta(k;s)) \sin(x \ \vartheta(k;s))
\end{equation}
and
\begin{equation}
	\vartheta(k;s)=\frac{k \ \pi}{s+1}.
\end{equation}
Call
\begin{equation}
	\rho_m(t)=\ketbra{M(t)}{M(t)}
\end{equation}
the density matrix of the machine at time $t$.\\
By taking the partial trace $Tr_{\mathcal{H}_{cursor}}(\rho_m(t))$ with respect to the cursor degrees of freedom, we get the density matrix $\rho_r(t)$ of the register:
\begin{equation} \label{eq:rhort}
	\rho_r(t)=\sum_{x=1}^s |c(t,x;s)|^2 \ \rho(x),
\end{equation}
where, for $1<x \leq s$,
\begin{equation}
	\rho(x)= U_{x-1}\ldots U_1 \rho(1) U_1^{-1}\ldots U_{x-1}^{-1}. 
\end{equation}
In order to trace out the register degrees of freedom, it is expedient to refer $\mathcal{K}_r$, at each time $t$, to the orthonormal basis $\ket{b_1(t)},\ket{b_2(t)},\ldots,\ket{b_d(t)}$ formed by the eigenvectors of $\rho_r(t)$:
\begin{equation}
	\rho_r(t) \ket{b_j(t)}= \lambda_j(t)\ket{b_j(t)},\ 1 \leq j \leq d.
\end{equation}
A simple computation shows, then, that the density matrix of the cursor is given by
\begin{equation}
	\rho_c(t)= \sum_j \lambda_j(t) \ketbra{d_j(t)}{d_j(t)}
\end{equation}
where the sum extends to the values of $1 \leq j \leq d$ such that $\lambda_j(t) >0$ and where, for such values of $j$, $\ket{d_j(t)}$ is given by
\begin{equation}
	\ket{d_j(t)}= \frac{1}{\sqrt{\lambda_j(t)}} \sum_{x=1}^s c(t,x;s) \braket{b_j(t)}{R(x)}\> \ket{C(x)}.
\end{equation}
As it is easy to check that $\braket{d_j(t)}{d_k(t)}=\delta_{j,k}$, the above representation of $\rho_c(t)$ shows that it has eigenvalues $\lambda_j(t)$ and, therefore, \emph{von Neumann entropy}
\begin{equation}
	S(\rho_c(t))=- \sum_j \lambda_j(t) \ln \lambda_j(t)= S(\rho_r(t)).
\end{equation}
The equality, in the particular case considered here, in which the initial state of the machine is the pure state $\ket{M(0)}= \ket{R(1)} \otimes \ket{C(1)}$, between the entropy $S(\rho_c(t))$ of the cursor and the entropy $S(\rho_r(t))$ of the register  (a well known property of a bipartite system in a pure state) is easily understood, in physical terms, by the insertion, in the expression for \ket{M(t)}, of the partition $I_r = \sum_{j=1}^d \ketbra{b_j(t)}{b_j(t)}$ of the identity in $\mathcal{K}_r$:
\begin{eqnarray} \label{eq:ent}
\ket{M(t)} & = & \sum_{j=1}^d \sum_{x=1}^s c(t,x;s) \braket{b_j(t)}{R(x)} \> \ket{b_j(t)} \otimes \ket{C(x)} =\nonumber \\
  & = & \sum_{j=1}^d \ket{b_j(t)} \otimes \sum_{x=1}^s c(t,x;s) \braket{b_j(t)}{R(x)} \> \ket{C(x)}= \nonumber \\
  & = & \sum_{j=1}^d \sqrt{\lambda_j(t)}\> \ket{b_j(t)} \otimes \ket{d_j(t)}.
\end{eqnarray}
The above expression (the \emph{Schmidt decomposition} of the state \ket{M(t)}) shows that if upon a measurement at time $t$, the register is found in state \ket{b_j(t)}, then the cursor collapses into the state  \ket{d_j(t)}, and vice versa.
\section*{An explicitly solvable example}
\begin{figure}[htbp]
	\centering
	\resizebox{7cm}{7cm}{
		\includegraphics{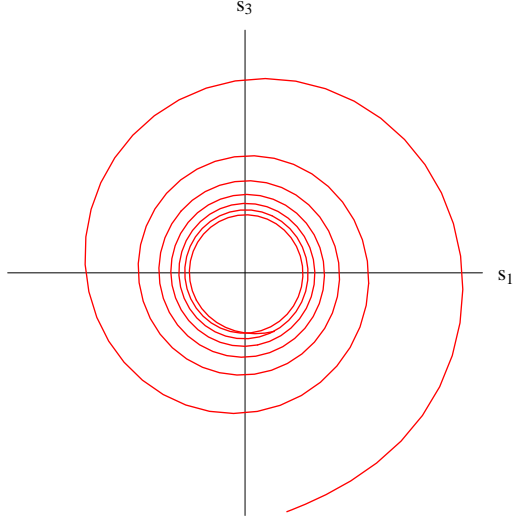}
		}
	\caption{A parametric plot of $\left (s_1(t),s_3(t)\right )$ for \mbox{$0 \leq t < s$}, $\lambda=1$. The choice $\mu=7,\;\chi= \arcsin(1/2^{\mu/2}),\; s=2^\mu +1,\; \alpha = -4 \chi,$ \mbox{$ \theta= \pi -2 \chi$} of the parameters is motivated by the connection with Grover's algorithm.}
	\label{fig:spirale}
\end{figure}

We focus our attention, in what follows, on the simplest non trivial case, in which $dim(\mathcal{H}_{register})=2$. We consider, namely, the simple case in which the register is a two level system or, equivalently, a spin $1/2$ system.\\
We indicate by  $\underline{\sigma}=(\sigma_1,\sigma_2,\sigma_3)$ the three components of such a spin in an assigned reference frame and by  $\underline{e}_1, \underline{e}_2,\underline{e}_3$ the versors of the three coordinate axes.\\
In the basis $\ket{\sigma_3= \pm 1}$, the density operator $\rho_r(t)$ will be represented by a matrix of the form
\begin{equation} \label{eq:rhor}
\rho_r(t)=\frac{1}{2} \left (
\begin{array}[pos]{c c}
1+s_3(t) & s_1(t)-i\>s_2(t)\\
s_1(t) + i\>s_2(t)  & 1-s_3(t)
\end{array}
\right )
\end{equation}
where
\begin{equation}
s_j(t)=Tr \left ( \rho_r(t) \cdot \sigma_j \right ),\;j=1,2,3.
\end{equation}
Equivalently stated, the \emph{Bloch representative} of the state $\rho_r(t)$ is given by the three-dimensional real vector
\begin{eqnarray}
\underline{s}(t)& = & s_1(t) \cdot \underline{e}_1+s_2(t)\cdot \underline{e}_2+s_3(t) \cdot \underline{e}_3= \nonumber \\
& = &\sum_{x=1}^s  \left | c(t,x;s)  \right | ^2 \; Tr(\rho(x)\cdot \underline{\sigma}).
\end{eqnarray}
We examine, here, the behaviour of $\underline{s}(t)$ in the simple example defined by the following additional conditions:
\begin{enumerate} 
\def\theenumi{\roman{enumi}}
\item  The initial state of the register is:
\begin{equation} \label{eq:stini}
\ket{R(1)}=\cos \left ( \frac{\theta}{2} \right ) \ket{\sigma_3=+1} + \sin \left(  \frac{\theta}{2} \right ) \ket{\sigma_3=-1},
\end{equation}
namely the eigenstate belonging to the eigenvalue $+1$ of $\underline{n}(1) \cdot \underline{\sigma}$, with
\begin{equation}
\underline{n}(1)= \underline{e}_1 \sin\theta + \underline{e}_3 \cos \theta. 
\end{equation}
\item Each of the unitary transformations $U_x$ is a rotation of a fixed angle $ \alpha $ around the axis $\underline{e}_2$
\begin{equation} \label{eq:stop}
U_1=U_2= \ldots = U_{s-1} = e^{-i \frac{\alpha}{2} \sigma_2}.
\end{equation} 
\end{enumerate}
We wish to remark that the above example captures the geometric aspects not only of such simple computational tasks as $NOT$ or $\sqrt{NOT}$ (viewed as rotations of an angle $\pi$ or $\pi/2$ respectively, decomposed into smaller steps of amplitude $\alpha$) but also of Grover's quantum search \cite{grover96}.\\
If the positive integer $\mu$ is the length of the marked binary word to be retrieved, set
\begin{equation}
\chi(\mu)=\arcsin(2^{-\frac{\mu}{2}})
\end{equation}
and
\begin{equation} \label{eq:eta}
\theta= \pi-2\>\chi(\mu).
\end{equation}
Then the state (\ref{eq:stini}) correctly describes the initial state $\ket{\iota}$ of the quantum search as having a component $2^{- \mu/2}$ in the direction of the target state, here indicated by $\ket{\omega}=\ket{\sigma_3=+1}$, and a component $\sqrt{1-2^{-\mu}}$ in the direction of the flat superposition, here indicated by $\ket{\sigma_3=-1}$, of the $2^\mu-1$ basis vectors orthogonal to the target state.
In this notations, if
\begin{equation} \label{eq:alpha}
\alpha=-4\>\chi(\mu),
\end{equation}
then the unitary transformation $\exp(-i \> \alpha\> \sigma_2/2)$ corresponds to the product $B\cdot A$ of the \emph{oracle} step
\begin{equation}
A=I_r-2 \> \ketbra{\omega}{\omega}
\end{equation}
and the \emph{estimation} step
\begin{equation}
B=2\>\ketbra{\iota}{\iota}-I_r
\end{equation}
where $I_r$ is the identity operator in $\mathcal{H}_{register}$.\\
It is having in mind the connection with Grover's algorithm that, for the sake of definiteness, in the examples that follow we are going to consider the one-parameter family of models, parametrized by the positive integers $\mu$, corresponding to the choice (\ref{eq:eta}) and (\ref{eq:alpha}) of the parameters $\theta$ and $\alpha$ and to the choice $s=2^\mu+1$ of the number of cursor sites, corresponding to the possibility of performing up to an exhaustive search.\\
In the example defined by (\ref{eq:stini}) and (\ref{eq:stop}) it is
\begin{equation}
\bra{R(x)}\> \underline{\sigma} \ket{R(x)} = \sin \left ( \theta+(x-1)\alpha \right)\> \underline{e}_1+\cos \left ( \theta+(x-1)\alpha \right)\> \underline{e}_3
\end{equation}
and, therefore,
\begin{equation}
\underline{s}(t) = \sum_{x=1}^s \left |c(t,x;s) \right | ^2  \left( \sin \left ( \theta+(x-1)\alpha \right)\> \underline{e}_1+\cos \left ( \theta+(x-1)\alpha \right)\> \underline{e}_3 \right ).
\end{equation}
Figure \ref{fig:spirale} presents a parametric plot of $(s_1(t),s_3(t))$ under the above assumptions.

\begin{figure}[htbp]
	\centering
		\includegraphics[width=7cm]{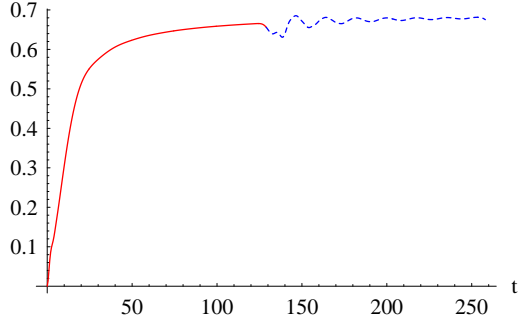}
	\caption{The von Neumann entropy of the register as a function of time, for the same model as in figure \ref{fig:spirale}, for $0 \leq t <2s$. The dashed part of the graph shows the effect of the cursor wave packet being reflected at the rightmost site $s$.}
	\label{fig:VN}
\end{figure}

It is convenient to describe the Bloch vector $\underline{s}(t)= s_1(t)\> \underline{e}_1+ s_3(t)\> \underline{e}_3$ in polar coordinates as
\begin{equation}
\begin{array}{lcr}
s_1(t)=r(t)\sin{\gamma(t)},
& &
s_3(t)=r(t)\cos{\gamma(t)}.
\end{array}
\end{equation}
The eigenvalues of $\rho_r(t)$ can then be written, in this notation, as
\begin{equation}
\begin{array}{lcr}
\lambda_1(t)=\frac{1}{2} (1+r(t)), 
& &
\lambda_2(t)=\frac{1}{2} (1-r(t))
\end{array}
\end{equation}
and the corresponding eigenvectors as
\begin{equation}
	\begin{array}{lcr}
		\ket{b_1(t)}= 
		\left (
				\begin{array}{c}
				\cos(\gamma(t)/2)\\
				\sin(\gamma(t)/2)
				\end{array}
		\right ),
		& &
		\ket{b_2(t)}= 
		\left (
				\begin{array}{c}
				-\sin(\gamma(t)/2)\\
				\cos(\gamma(t)/2)
				\end{array}
		\right ).
	\end{array}
\end{equation}
The two corresponding states of the cursor, in the sense of (\ref{eq:ent}), are then:
\begin{eqnarray}
\ket{d_1(t)}=\frac{1}{\sqrt{\lambda_1(t)}} \sum_{x=1}^s c(t,x;s)\> \cos \left(  ( \theta + (x-1)\alpha - \gamma(t)/2)\right) \> \ket{C(x)}; \\
\ket{d_2(t)}=\frac{1}{\sqrt{\lambda_2(t)}} \sum_{x=1}^s c(t,x;s)\> \sin \left(  ( \theta + (x-1)\alpha - \gamma(t)/2)\right) \> \ket{C(x)}.
\end{eqnarray}
The von Neumann entropy $S\left( \rho_r(t) \right)$ is therefore
\begin{equation} \label{eq:vne}
S\left( \rho_r(t) \right) = - \frac{1+r(t)}{2} \ln \frac{1+r(t)}{2} - \frac{1-r(t)}{2} \ln \frac{1-r(t)}{2}.
\end{equation}
An example of its behaviour is shown in figure \ref{fig:VN}.
\begin{figure}[htbp]
	\centering
		\includegraphics[width=7cm]{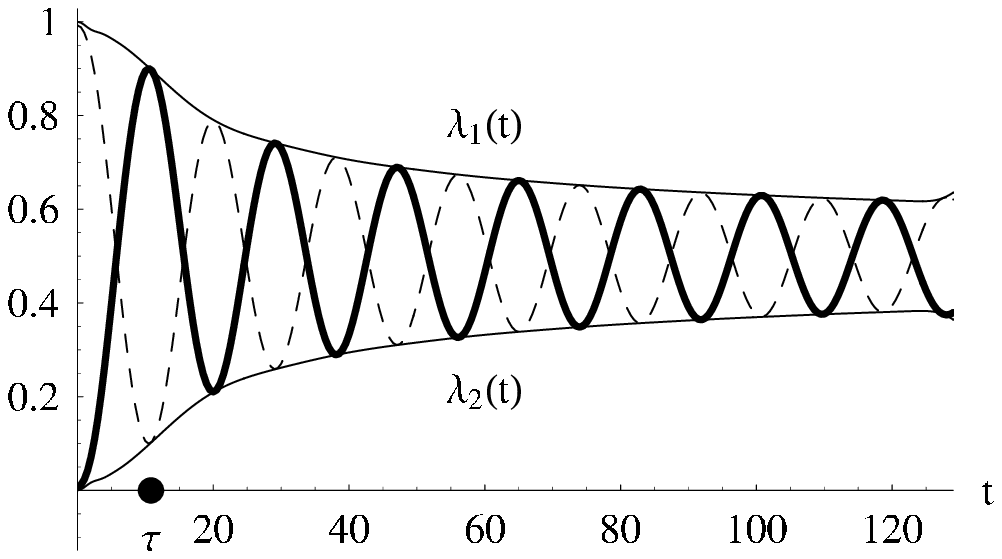}
	\caption{The thick line is a graph, for $0 \leq t <s$ of $Tr(\rho_r(t)(I+\sigma_3)/2)$, the probability of observing the ``target'' state $\ket{\omega}=\ket{\sigma_3=+1}=\ket{B_1}$ in the example of Grover's algorithm. We have evidenced the instant $\tau=O(2^{\mu/2})$ at which this probability reaches its maximum. The dashed line is a graph of \mbox{$Tr(\rho_r(t)(I-\sigma_3)/2)$}, the probability of observing the ``undesired'' output\mbox{ $\ket{\sigma_3=-1}=\ket{B_2}$}.}
	\label{fig:limiti}
\end{figure}
It is to be stressed that, as $\lambda_1(t)>\lambda_2(t)$, at each time $t$ the projector $\ketbra{b_1(t)}{b_1(t)}$ is, among the projectors on the state space of the register, the one having in the state $\rho_r(t)$ the greatest probability of assuming, under measurement, the value $+1$.\\
Stated otherwise, if the desired output of the computation is a given state $\ket{B_1}$ (the eigenstate corresponding to the eigenvalue $+1$ of the component of $\underline{\sigma}$ along an assigned direction in the $\underline{e}_1,\underline{e}_3$ plane) the optimal choice of the time $t$ at which to read the output is such that $\ket{b_1(t)}=\ket{B_1}$.\\
What figures \ref{fig:spirale} and \ref{fig:VN} show is that for no choice of $t > 0$ is the probability of finding the ``target'' output $\ket{B_1}$ equal to 1: it is bounded above by $\lambda_1(t)$. There is always, as shown in figure \ref{fig:limiti}, a non vanishing probability (bounded below by  $\lambda_2(t)$) of finding the register in the orthogonal, ``undesired'',  state \ket{B_2}.
\section*{Reading the register}
With reference, for definiteness, to the example of figure 3, call $\tau$ the instant of time at which the probability  $Tr(\rho_r(t)\ \ketbra{\omega}{\omega})$ reaches its first  and absolute maximum. We recall that, in the above example, the target state  $\ket{\omega}$ is taken to be the ``up'' state \ket{\sigma_3=+1} of the register.\\
The whole point of the analysis of the previous section is that $\lambda_1(\tau)$ is \emph{strictly} smaller than $1$ . This amounts, in turn, to a deficit  $1-\lambda_1(\tau)$  in the probability of finding the target state. This deficit is not, in itself, a strong limitation in a quantum search algorithm, because we can in principle identify the right target through a majority vote among a ``gas'' of a large number $N$  of machines. The trouble is that  if we want to use the same machines once more, we need to purify the ``gas'' of registers from the fraction $\lambda_2(\tau)$  of them which have collapsed into the wrong state: standard thermodynamic reasoning \cite{landauer61}  shows then that this requires the removal from the gas, supposing a heat reservoir at temperature $T$ is available, of an amount of heat of $N k_B T S(\rho_r(\tau))$, $k_B$ being Boltzmann's constant.\\
We wish, in this section, to supplement the above considerations with an explicit description of the post-measurement state of the machine, showing, in particular, the effect \emph{onto the clock} of the act of reading the register.\\[5pt]
Suppose that at the optimally chosen instant $\tau$, at which it is $\gamma(\tau)=0$, while the machine is in the state  \ket{M(\tau)},  a measurement of the projector $(I_r+\sigma_3)/2$ is performed.\\
If the measurement gives the result $1$, then the state \ket{M(\tau)} collapses to
\begin{equation}\label{eq:M1}
	\ket{M_1(\tau)}=\ket{\sigma_3=+1} \otimes \frac{1}{\sqrt{\lambda_1(\tau)}} \sum_{x=1}^s c(\tau,x;s)\cos((\theta+(x-1)\alpha)/2)\ket{C(x)}.
\end{equation}
If, instead, the measurement gives the result $0$, then the state \ket{M(\tau)} collapses to
\begin{equation}\label{eq:M2}
	\ket{M_2(\tau)}=\ket{\sigma_3=-1} \otimes \frac{1}{\sqrt{\lambda_2(\tau)}} \sum_{x=1}^s c(\tau,x;s)\sin((\theta+(x-1)\alpha)/2)\ket{C(x)}.
\end{equation}
Figures 4.a and 4.b show the probability distributions
\begin{eqnarray}
P_1(x,\tau)& = & \left| \left(c(\tau,x;s) \cos((\theta+(x-1) \alpha)/2)\right) \right|^2 / \lambda_1(\tau)\\
P_2(x,\tau)& = & \left| \left(c(\tau,x;s) \sin((\theta+(x-1) \alpha)/2)\right) \right|^2 / \lambda_2(\tau)
\end{eqnarray}
of the observable $Q$ (position of the cursor)  in the states  \ket{M_1(\tau)} and \ket{M_2(\tau)}, respectively.
\begin{figure}[h]
	\centering
		\includegraphics[width=55mm]{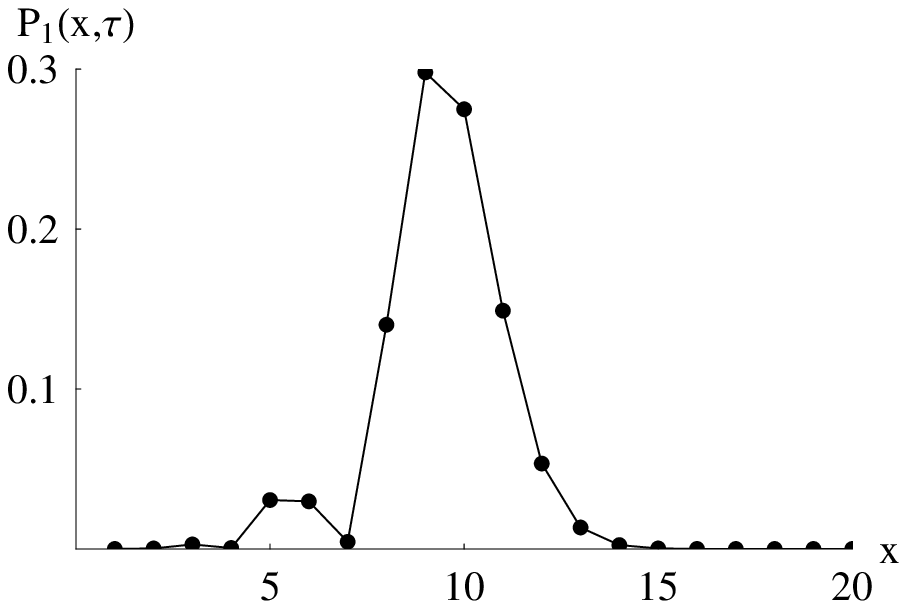}
		\hspace{0.5cm}
		\includegraphics[width=55mm]{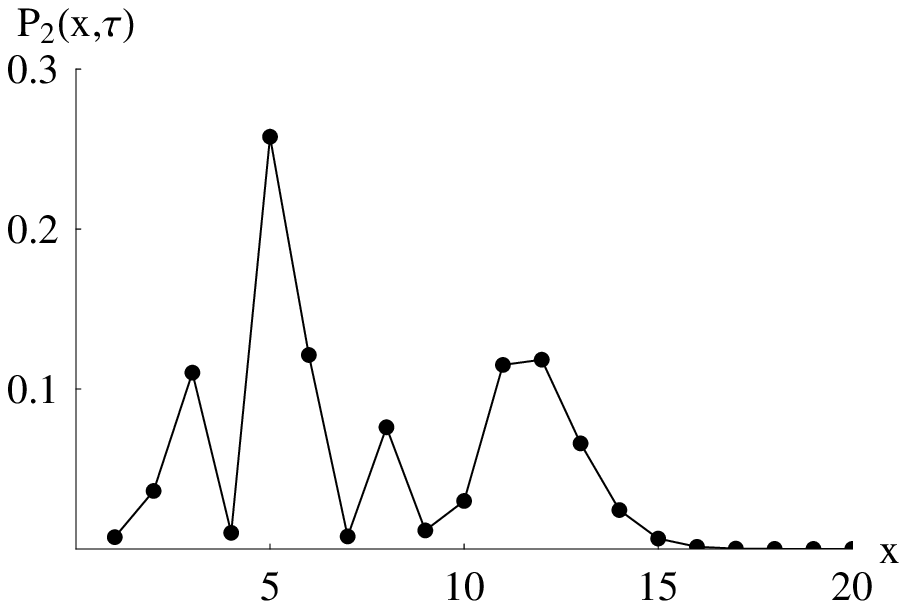}
		\begin{picture}(230,10)
			\put(20,0){(a)}
			\put(200,0){(b)}
		\end{picture}
	\caption{Figures (a) and (b) represent, for the same choice of parameters as in figure \ref{fig:spirale}, respectively the probabilities $P_1(x,\tau)$ and   $P_2(x,\tau)$ as  functions of $x$.}
	\label{fig:VNSh}
\end{figure}
Figures 5.a and 5.b show the probability distributions of the observable $H$ (energy of the machine)  in the states \ket{M_1(\tau)} and \ket{M_2(\tau)}, respectively.
\begin{figure}[htbp]
	\centering
		\includegraphics[width=55mm]{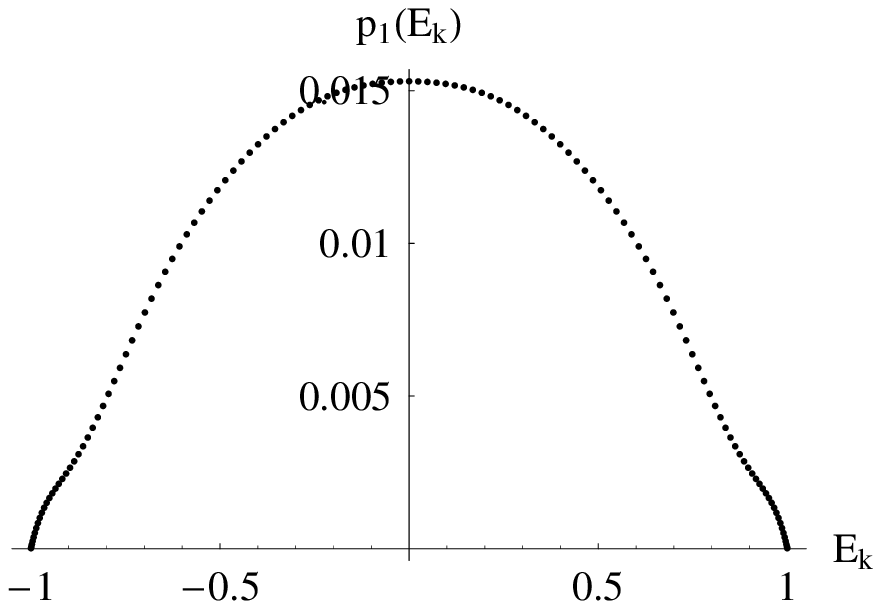}
		\hspace{0.5cm}
		\includegraphics[width=55mm]{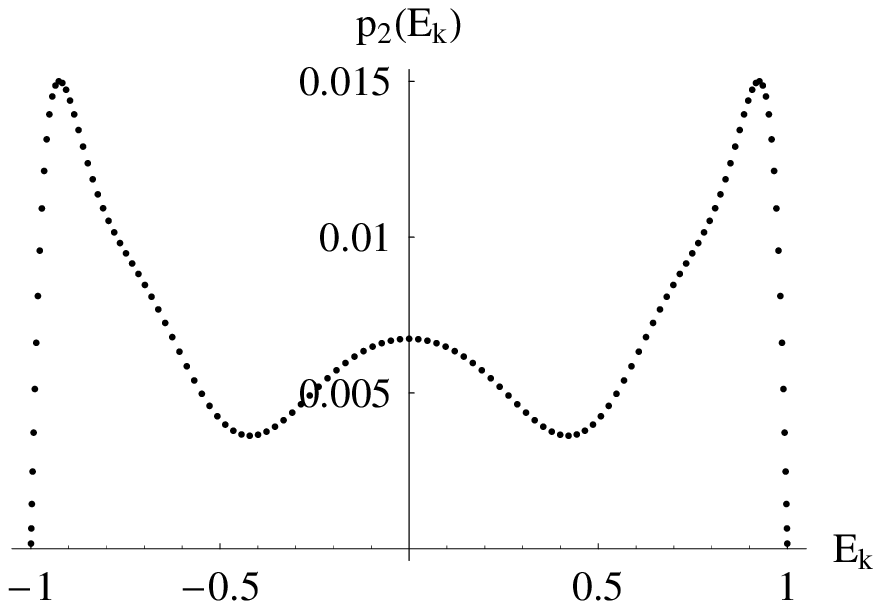}
		\begin{picture}(230,10)
			\put(20,0){(a)}
			\put(200,0){(b)}
		\end{picture}
		\caption{Figures (a) and (b) represent the probability distribution $p_1(E_k)$ and $p_2(E_k)$ of the energy $H$ in the state \ket{M_1(\tau)} and \ket{M_2(\tau)} respectively.}
	\label{fig:distribuzioneenergia}
\end{figure}
The two energy distributions of figures 5 are easily derived from the fact that the Hamiltonian $H$ defined in (4) has the eigenvalues
\begin{equation}\label{eq:eigenvaluesH}
	E_k= -\lambda\; \cos(\vartheta(k;s)),\ k=1,\ldots,s;
\end{equation}
each doubly degenerate, an orthonormal basis in the eigenspace belonging to the eigenvalue $E_k$ being given, for instance, by the two eigenvectors
\begin{equation}
	\ket{E_k;\sigma_2=\pm 1}= \ket{\sigma_2=\pm 1} \otimes \sum_{x=1}^s v_k(x) \exp(\mp i \alpha (x-1)/2)\ket{C(x)},
\end{equation}
where
\begin{equation}
	v_k(x)=\sqrt \frac{2}{s+1} \sin(x\;\vartheta(k;s)).
\end{equation}
This leads to the explicit expressions
\begin{equation}
	p_j(E_k)=\sum_{\eta=\pm 1} \left|\braket{M_j(\tau)}{E_k;\sigma_2=\eta} \right|^2, \ j=1,2.
\end{equation}
Figures 4 and 5 show that a collection of identically prepared and independently evolving machines becomes in fact, under the operation of reading the register at time $\tau$, a mixture of two distinct ``molecular'' species, ``1'' (present in a concentration $\lambda_1(\tau)$ ),  and ``2'' (present in a concentration $\lambda_2(\tau)$). In each of these two molecular species, the same ``atomic'' constituents have arranged themselves in a different geometrical shape (figures 4), with a different orientation of the register spin (equations (\ref{eq:M1}) and (\ref{eq:M2})),  because of a different energy distribution (figures 5).\\
Comparison with the distribution of $H$ in the pre-measurement state \ket{M(\tau)}, given in figure 6, shows that the presence of the impurities of type ``2'' is due to unusually intense exchanges of energy between the machine and the reading (measurement) apparatus.
\begin{figure}[htbp]
	\centering
		\includegraphics[width=7cm]{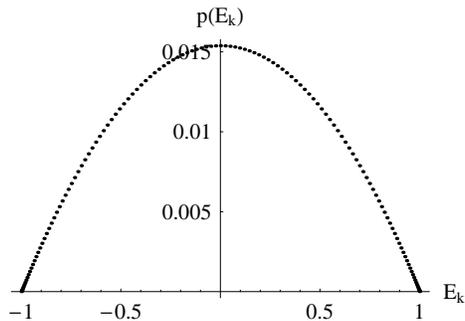}
	\caption{In the state \ket{M(\tau)} the probability distribution of $H$ is given by $p(E_k)=(v_k(x))^2$.}
	\label{fig:spettro}
\end{figure}

\section*{Discussion and outlook}
We don't claim that (quantum) computation cannot be made reversible. We have simply pointed out one aspect in which a reversible machine is an idealization; this idealization amounts to neglecting the back-reaction (figures \ref{fig:distribuzioneenergia}) of the clocked subsystem onto the clock.\\
In very  concrete physical terms, from the dispersion relation $E(p)= -\cos(p)$ (see equation (\ref{eq:eigenvaluesH})) it is immediate to conclude that the distortion of the energy spectrum depicted in figure \ref{fig:distribuzioneenergia}.b corresponds to a decrease in the speed $v=dE(p)/dp$ of the cursor. This recoil effect could of course be neglected if the clock consisted of, say, $10^{23}$ atoms (we refer the interested reader to the huge literature on the limitations posed by quantum mechanics to the measurement and/or operational definition of space-time distances \cite{gambini04}\cite{salecker58}), but might be of relevance for a machine scaled down to a molecular chain evolving under the sole effect of its initial condition not being an eigenstate of the Hamiltonian.\\
Stated in equivalent terms, there is nothing wrong in the assumption of starting the computation in a pure state of the register: we have pointed out, however, that realizing this initial condition has a cost $N k_B T S(\rho_r(\tau))$ in terms of energy to be dumped into the environment in order to get rid of the entropy generated (for the simple fact that the two recoil patterns of figures \ref{fig:distribuzioneenergia}.a and \ref{fig:distribuzioneenergia}.b have \emph{both}  strictly positive probability) in the previous run of the machine.\\
The toy model corresponding to the choice  (\ref{eq:stop}) has allowed us to show in an explicit quantitative form the decoherence induced by the coupling with the timing apparatus, appearing through the build-up of entropy in the state of the register subsystem.\\
Beyond the details of the model considered, it is to be stressed that such a build-up is a general consequence of the fact that the coefficients $c(t,x;s)$ appearing in (\ref{eq:rhort}) are determined by the discretized version of the free Schr\"odinger equation 
\begin{equation}\label{eq:schroedinger}
	i \frac{d}{dt} c(t,x;s)= - \frac{\lambda}{2}\left(c(t,x-1;s)+c(t,x+1;s)\right),
\end{equation}
leading to the well known phenomenon of wave packet spreading (quadratic increase in time of the variance of $Q$).\\
We are well aware that the explicit model discussed in the previous sections is, under many respects, far from being optimized from the point of view of minimizing the probability deficit  $1-\lambda_1(\tau)$.\\
The initial condition (\ref{eq:condiniziali}), for instance, corresponds to the classical intuition of initially placing the clock in a sharply defined position \ket{C(1)}. It will be worth studying the effect of a better choice of initial conditions, with the probability amplitude of the cursor spread on an initial extended region; in much the same spirit, we recall the analysis leading in \cite{apolloni02} to the proposal of supplementing the active part of the cursor subsystem (the collection of sites $x$ for which $U_x$ is different from the identity) with an extended inactive part, having the effect of temporarily lowering the entropy of the register subsystem. The problem of optimally investing, in the above two ways (to the left and to the right of the active region), an assigned amount of space resources in order to minimize at a selected time $\tau$ the deficit $1-\lambda_1(\tau)$ in the probability of finding the target state poses itself as a natural question in this context.\\
A related problem is that of providing a stability analysis of motion under position dependent coupling constants $\lambda(x)$ in (\ref{eq:schroedinger}).\\ 
Under two more respects, in studying a model of the class defined by (\ref{eq:hamiltonian}), we have not fully exploited the potentialities of Feynman's approach:
\begin{itemize}
\item in Feynman's full model  the quantum walk performed by the cursor is by no means restricted to a linear graph: the use of conditional jumps allows, in fact, to explore much more interesting planar graphs (we refer to \cite{defa04} for a quantitative study of these more general systems);
\item  the notation adopted in (\ref{eq:hamiltonian}) does not give a full account of the original intuition (better described in terms of creation and annihilation operators $\tau_\pm(x)$ ) of a single particle in a quantum lattice gas  jumping between nearest neighbor sites.
\end{itemize}
\vspace{-0.4cm}
Having written the Hamiltonian (\ref{eq:hamiltonian}) as
\begin{equation} \label{eq:hamiltonian1}
	H= -\frac{\lambda}{2} \sum_{x=1}^{s-1} U_x \otimes \tau_+(x+1)\tau_-(x)+ U^{-1}_x \otimes \tau_+(x)\tau_-(x+1),
\end{equation}
the idea emerges quite naturally of studying the evolution under (\ref{eq:hamiltonian1}) of  \emph{many} particles.\\
Possible applications of this proposal of a ``multi-hand quantum clock'', its use in the repeated application of a given transformation, the steering of the particles along different branches of the graph in order to act in parallel on distinct parts of the register, the solution of  conflicts in the application of non commuting primitives to a same part of the register deserve, we think, further research.

\section*{Acknowledgements}
We wish to thank prof. Alberto Bertoni for his constant encouragement and constructive criticism.

\bibliography{biblio}
\bibliographystyle{abbrv}
\end{document}